 \documentclass[manuscript]{aastex}

\usepackage{emulateapj5}

\newcommand{\sgr}{Sgr~A$^\ast~$}
\newcommand{\rs}{$R_{\rm S}~$}
\newcommand{\dmn}{\dot{m}_{-6}}
\newcommand{\dm}{\dot{m}}

\shorttitle{Black Hole Plerion}
\shortauthors{Atoyan \& Dermer}

\received{}

\begin{document}

\title{TeV Emission from the Galactic Center 
Black-Hole Plerion }

\author{Armen Atoyan$^1$ and Charles D. Dermer$^2$} 
\affil{$^1\,$CRM, Universit\'e de Montr\'eal, C.P. 6128, 
Montr\'eal, Canada H3C 3J7;~~{atoyan@crm.umontreal.ca}}

\affil{$^2\,$E\,O.\,Hulburt Center for Space Research, Code 7653,
Naval Research Laboratory,\\ Washington, DC 20375-5352;~~
dermer@gamma.nrl.navy.mil}

\begin{abstract}
The HESS collaboration recently reported highly significant
detection of TeV $\gamma$-rays coincident with Sgr~A$^\ast$
\citep{HESS}.  In the context of other Galactic Center (GC)
observations, this points to the following scenario: In 
the extreme advection-dominated accretion flow (ADAF) 
regime of the GC black hole (BH), synchrotron radio/sub-mm 
emission of $\sim 100 \,\rm MeV$ electrons emanates from 
an inefficiently radiating turbulent magnetized corona 
within 20 $R_{\rm S}$ (Schwarzschild radii) of the
GCBH.  These electrons are accelerated through 
second-order Fermi processes by MHD turbulence, 
as suggested by Liu et al.\ (2004).
Closer to the innermost stable orbit of the ADAF, 
instabilities and shocks within the flow inject 
power-law electrons through first-order Fermi acceleration 
to make synchrotron X-ray flares observed with Chandra, 
XMM, and INTEGRAL. A subrelativistic MHD wind subtending a
$\sim 1\,\rm sr$ cone with power 
$\gtrsim 10^{37}\,\rm erg\,s^{-1}$ is
driven by the ADAF from the vicinity of the GCBH.   
As in pulsar powered plerions, electrons are
accelerated at the wind termination
shock, at $\gtrsim 10^{16.5}\,$cm from the GCBH,  and 
Compton-scatter the ADAF and the far infra-red (FIR) 
dust radiation 
to TeV energies. The synchrotron radiation of these
electrons forms the quiescent X-ray source resolved by
Chandra.  The radio counterpart of this TeV/X-ray plerion, 
formed when the injected electrons cool on timescales 
of $\gtrsim 10^4\,\rm yrs$,
could explain the origin of nonthermal radio emission in the
{\it pc}-scale bar of the radio nebula Sgr A West.
\end{abstract}


\maketitle

\vspace{3mm}

\section{Introduction and Observations}

HESS measurements of TeV fluxes from Sgr A$^\ast$ 
\citep{HESS} confirm, with high significance, earlier 
reports by the Whipple \citep{Whipple} and 
CANGAROO \citep{CANGAROO} collaborations
(albeit at higher flux levels than HESS) 
about TeV emission from the Galactic Center. This result
adds 
important new information about the massive ($M_{GC}\approx 
4\times 10^6 M_\odot$; \citet{ghe03}) BH at the center
of the Galaxy.  When combined with the wealth of
observational data in the radio to X-ray wavebands  
\citep{mf01}, the TeV $\gamma$-ray data point to a 
scenario involving an inner,
inefficiently radiating magnetized corona, i.e.\ 
the ADAF, and a
subrelativistic outflow of particles and field from the 
ADAF. As in pulsar plerions like Crab Nebula 
(see \citet{kc84}),
the wind terminates at a shock, at $R \gtrsim 3\times 10^{16}\,\rm cm$
from the GCBH, to create a black-hole powered `plerion' where the TeV
radiation is produced.

The VLBI size of the radio source at the location of \sgr is
$110 (\pm 60)\,\mu$as at $\nu = 215 \,\rm ~GHz$ and $190 (\pm
30)\,\mu$as at $\nu = 87 \,\rm GHz$ \citep{vlbi}.
Using VLBA, \citet{bow04} resolve  \sgr  at 42 GHz to have 
an  angular size of $240 (\pm  10)\,\rm \mu$, or 
$R_{rad} \simeq 24 \,R_{\rm S}$ for 
$M_{GC} = 4\times 10^{6} M_\odot$ black hole and a GC 
distance $d= 8\,\rm kpc$,  
and find that the size increases
with wavelength. The radio emission of \sgr is rather stable,
with total luminosity $L_{36} = L_{rad}/ 10^{36} \,\rm erg 
\, s^{-1}\simeq 1$ and spectral flux $S_\nu \propto 
\nu^{0.3}$ that sharply falls off above 
$\nu_0 \sim 10^{12}\,\rm Hz$ \citep{Z95,fal98}.

The near-infrared (NIR) flares 
show factors-of-2 flux variations on timescales of 
$\sim 20\,$min and reach powers of 
$\approx 10^{35}\,\rm erg\,s^{-1}$ \citep{gen03}.  The
spectra rise in $\nu S_\nu$ diagrams, 
and show powers $\approx 10^{35}\,\rm erg\,s^{-1}$, 
similar with the ({\it nonsimultaneously}) detected
X-ray flares \citep{bag01,por03}. The 200\,s X-ray 
variability timescale limits the engine size to $r
=R/R_{\rm S} \lesssim 5$. The unabsorbed quiescent flux
detected by Chandra observatory from \sgr is 
$L_{X.qui}(2$--$10\, \rm keV) \approx 2.4 \times 
10^{33}\,\rm erg\, s^{-1}$, and has angular extent 
$\theta_{\rm FWHM}\approx 1.4^{\prime\prime}$ 
\citep{Chandra}, corresponding to source size 
$R_X\approx 8.3\times 10^{16}\,\rm cm$.

The TeV spectral fluxes in two HESS campaigns in 
June/July and July/August of 2003 are almost exactly the 
same, extending with photon
index $\approx \! 2.1$-2.3 from $160\,$GeV to 10 TeV, 
with total power 
$L_\gamma  \approx 10^{35}\,\rm erg\, s^{-1}$.
There is no sign of flux variation within the statistics 
available in
each of these sets \citep{HESS}.  Given the high statistical
significance of the signal in both data sets, 
these results suggest a quiescent or quasi-stationary TeV 
source coincident within 
$\leq \! 1.5^\prime$ of Sgr A$^\ast$.

We propose a model for Sgr A$^\ast$ where the 
quasi-stationary radio
and flaring X-ray/NIR emissions are synchrotron radiation 
from the ADAF. The MHD wind from the ADAF powers a 
black-hole plerion where the
quiescent X-ray and TeV emissions are produced by electrons
accelerated at the wind shock.  The model explains the existing
multi-wavelength data, and makes predictions that can be checked by
various X-ray and gamma-ray detectors.

\section{Parameters of the ADAF}

The dimensionless mass-accretion rate (or Eddington ratio)
$\dot{m}=\eta_{BH}\dot{M} c^2/L_{Edd}$ is defined in units of the
Eddington luminosity $L_{Edd}$, where $\eta_{BH} \simeq 0.1$ is the
maximum efficiency of gravitational-to-photon energy conversion by the
BH. In the ADAF model \citep{ny95,esi98}, thermal radiation of the
accretion plasma at the level $L_{T} = \dot{m}^2 L_{Edd}/\dot{m}_\ast$
is predicted when $\dot{m}\leq \dot{m}_\ast \sim 0.1$. The rate
$\dot{m} \simeq 1.5 \times 10^{-5}$ is obtained if one formally
equates $L_{T}$ with the observed $ L_{rad} 
\simeq 10^{36}\,\rm erg\, s^{-1}$. 

The compact radio emission cannot originate from an optically-thick 
 accretion disk  
which would  only be allowed at $R \gg  R_{rad} $, 
or from a much hotter
ADAF plasma at smaller scales, but is compatible with a synchrotron origin
\citep{lpm04}.  The thermal output of the
ADAF would be peaked mostly in the hard X-ray/soft gamma-ray domain,
where the quiescent luminosity is at least 2 orders of magnitude
smaller than $L_{rad}$.  This results in the accretion rate estimate
$\dmn = \dm /10^{-6} \lesssim 10^{-6} $ consistent with ADAF model and
\sgr observations.  Even for $\dm$ this small, $\lesssim 10^{39}\,\rm
erg\, s^{-1} $ of accretion power is still either advected into the BH
or escapes as an outflow.

Adopting GCBH mass $3\times 10^6 \,M_{\odot}$, the ADAF magnetic field
at radii $r= R/R_{\rm S}$ is \citep{ny95}
\begin{equation}
B (r)\approx 370 \alpha^{-1/2} \lambda^{1/2} \dmn^{1/2} r^{-5/4}
\; \rm G\, ,
\end{equation}
where $\lambda =1-\beta$ is the ratio of the magnetic to the total
pressure in the accreting gas and $\alpha$ is the viscosity parameter.
Using $\alpha \sim 0.1 $ and $\lambda
\sim 0.15$ \citep{qn99}, the magnetic field 
 $B \sim 10$ G in the ADAF at $r \lesssim
20$ is found. For these fields, the Lorentz factor of
electrons producing synchrotron radiation peaked at $\nu_0 \gtrsim
10^{12}\,\rm Hz$ is $\gamma_{0}\sim 200$, and the ADAF is still
transparent to self-absorption for $\gtrsim \! \rm GHz$ 
radiation.

A stochastic acceleration model for the radio-mm emission from
\sgr \citep{lpm04} can be quantified using the second-order
Fermi acceleration theory of \citet{dml96}. The magnetic field is
defined through the relation $B^2/8\pi = \epsilon_B \eta_{BH}\dot M
c^2/(4\pi R^2 c)$, implying $B \approx$ $ 30
\epsilon_B^{1/2}L_{36}^{1/4}$ G for a region of size $20 R_{\rm S}$.
Equating the acceleration rate of electrons by whistler turbulence
with the synchrotron loss rate gives a characteristic electron
Lorentz factor
$\gamma_{0} \cong 5\;(\zeta \epsilon_B)^{1/3} L_{36}^{1/6}
\tau_{\rm T}^{-11/18}$,
assuming a Kolmogorov spectrum for the turbulence. Here $\zeta$ is
the fractional whistler turbulent energy density compared to
$B^2/8\pi$, and $\tau_{\rm T}$ is the Thomson depth of the ADAF,
assumed to be composed of e-p plasma. Values of $B = 10$ G and
$\gamma_{0} \cong 200$ are implied for the ADAF when $\zeta \approx
0.1$, $\epsilon_B \approx 0.1$ and $\tau_{\rm T} \approx 2\times
10^{-4}$.

\section{Black Hole Plerion}

A Compton (i.e., leptonic) origin for non-variable TeV flux detected
with HESS implies that the synchrotron counterpart of that flux is the
quiescent component of the X-ray flux.  The magnetic-field energy
density in the source should then be $u_{B}=B^2/8\pi < 0.1 \, u_{rad}$,
where
\begin{equation} 
u_{rad}= \frac{L_{rad}}{4\pi c R^{2}} = 
2.65\times 10^{-8} {L_{36}\over  R_{16}} \,\rm erg \,cm^{-3}\,
\end{equation}
is the energy density of the target photons at $R=R_{16} \, 10^{16}
 \,\rm cm$.  Inside the radio source, at $r \lesssim 20$, the
 radiation energy density saturates at the level $u_{rad}\simeq
 L_{rad}/2\pi c R_{rad}^{2}$. Thus, the origin of TeV flux at those
 small distances would require magnetic fields $B \sim 0.1 \,\rm G$,
 far below the ADAF model values.

In the ADIOS (``advection-dominated inflow-outflow 
solution'' \citet{bb99}) extension of the ADAF model, 
the outflow may carry a
significant fraction of the generated kinetic energy of 
the plasma
from the BH vicinity to large distances. We assume a wind of
magnetized accretion plasmas with 
total power $L_{wind}= L_{w.37}\,10^{37}\,\rm erg\, s^{-1}$ 
propagating in a two-sided cone with
opening angle $\Omega$ at speed $v_{w} \! \lesssim \! c$; for
calculations we use $v_{w} \! =\! c/2$.  
This wind terminates at a subrelativistic shock at 
$R_{shock} \leq R_{X}$ from the GCBH. The shock accelerates 
electrons to energies $E_{max} =
\gamma_{max} m_e c^2 \gg 10 \,\rm TeV$ to produce the 
observed TeV $\gamma$-rays by Compton upscattering 
of sub-mm ($\nu_0 \sim 10^{12}\,\rm Hz$) photons from 
the ADAF. Another important photon target is
the emission from the cold ($\simeq 100\,\rm K$) dust 
ring of Sgr A
West, with total luminosity $\simeq 5\times 10^{6} L_{\odot}$ in the
central arcmin (few {\it pc}) region around \sgr
\citep{dust}. This radiation has a spectral peak at $\simeq 10^{13}
\,\rm Hz$ and estimated energy density $\simeq 2.4 \times 10^{3}\,\rm eV \,
cm^{-3}$. It becomes the main contributor to the $\lesssim 100 \,\rm GeV$
Compton flux from the plerion at 
$R\gtrsim 10^{17}\,\rm cm$ from Sgr A$^\ast$.  Synchrotron
radiation from these same multi-TeV electrons at $R\leq 10^{17}\,\rm cm$ 
produces the quiescent X-ray flux from Sgr A$^\ast$.

Chandra observations \citep{Chandra} reveal diffuse thermal X-rays
with luminosity $2.4\times 10^{34}\,\rm erg \, s^{-1}$ from the
central parsec region of Sgr A$^\ast$, produced by $k T \simeq
1.3\,\rm keV$ plasma with $n_e =26/\eta_{f}^{1/2}\,\rm cm^{-3}$
density, where $\eta_f =\eta_{-1}/10^{-1} $ is the volume filling
factor. Equating the gas pressure $n_e kT$ in that region with the
energy density of the wind gives the shock distance
\begin{equation} R_{shock} \simeq 3.1 \times 10^{16}\, L_{w.37}^{1/2}
\Omega_{w}^{-1/2}
\eta_{-1}^{1/4}
  \,\rm cm.
\end{equation}
The TeV luminosity $L_\gamma \simeq 10^{35} \,\rm erg \,s^{-1}$
implies a total electron acceleration power $L_e \gtrsim 10^{36}\,\rm
erg \, s^{-1}$. Assuming that the efficiency of electron acceleration
$L_e /L_{wind} \lesssim 30\%$, and letting the rest go to thermal
plasma and nonthermal protons (unlike in pulsar winds where
relativistic electrons dominate), we arrive at $L_{w.37} \gtrsim
0.3\,$.  Note that Eq.(3) predicts that the maximum power of the
outflow sustainable at distances $R_{shock} \leq R_{X}$ is $L_{w.37}
\lesssim 10$.

 The mean fluid speed $v_f$ in the TeV plerion formed 
downstream of the shock can be derived from $(R_X-R_{shock}) \simeq
v_f \times t_{C}(\gamma_X)$, where
\begin{equation}
t_{C}(\gamma_X)\simeq 4\times 10^7 \, { R_{16}^{2} \over L_{36}}\,  
\left( \frac{\epsilon_X}{1\,\rm keV}\right)^{-1/2} 
\; \rm s 
\end{equation}
is the Compton cooling time of electrons producing synchrotron X-rays
with energy $\epsilon_X$ for the estimated magnetic
field $B_{pler}\sim 100\,\rm \mu G$ in the GCBH plerion
at scales $R\lesssim R_X$.  Taking
$R_{16} = 5$ for the median distance of the TeV plerion, 
we find $v_f\gtrsim 500 \,\rm km/s$ at those distances.

Since relativistic electrons of all energies convect with 
the fluid at the same speed, the TeV plerion model predicts
shrinking of the quiescent X-ray source size at higher 
energies.  Note
also that the opening angle of the outflow should be 
$\Omega \sim 1$ rather than $\Omega \ll 1$, since otherwise 
$R_{shock} $ would exceed
$R_X$.  This indicates that the energy outflow  is in
the form of a wind rather than well-collimated jet.

\section{Spectral Modeling}

Fig.\ 1 shows numerical calculations of the quiescent 
radiation components of Sgr~A$^*$. 
These include synchrotron and SSC emission 
from the magnetized ADAF within $r_{rad}\approx 20$ of the 
GCBH ({\it dashed curves}), 
X-ray and TeV emission from the compact plerion 
at scales 
$R\sim (3$ -- $10)\times 10^{16}\,\rm cm$ ({\it solid curves}), 
and the emission from the larger plerion ({\it dot-dashed curves})
inflated to {\it pc} scales in the process of convective 
(and possibly also diffusive)
propagation of the accelerated electrons on timescales $\lesssim
10^4\,\rm yr$. Magnetic field in the latter,
 $B_2 = 140\,\rm \mu G$, is assumed higher than 
$B_{1}=90\,\rm \mu G$ in the TeV plerion. 
This is in agreement with the expected increase of the
magnetic field downstream of the shock in typical pulsar plerions 
\citep{kc84}, and is explained 
by deceleration and compression of the plasma.
But it is also possible that $B_2 < B_1$, in which case  a smaller 
synchrotron flux from the 
$pc$-scale plerion than shown in Fig.~1 is predicted .

The inset in Fig.\ 1 shows the nonthermal electron
distributions. Besides synchrotron and
Compton losses, calculations also include 
Coulomb and bremsstrahlung losses in a medium with
$n_{gas} = 10^3 \,\rm cm^{-3}$ characteristic for 
central $pc$ 
regions of Sgr~A$^*$. Electrons with a power-law injection
index $\alpha_{pler}=2.2$ and an exponential cutoff above 
$E_{max}=50\,\rm TeV$ are injected into 2-sided nebula with total
power $L_{e} = 6.5\times 10^{36}\,\rm erg \, s^{-1}$. These 
electrons
radiatively cool downstream of the shock and propagate to distances
$\geq 10^{17}\,\rm cm$ on the timescale $t_{esc} = 50\,$yr, 
during which
most of the injected energy of the multi-TeV electrons 
(producing TeV and X-ray fluxes observed) has been already lost.

The radio emission is produced 
at $r\lesssim 20 $ in $B\approx 10\,\rm G$ field 
by a relativistic Maxwellian electron
distribution $\propto \gamma^2 e^{-\gamma/\gamma_0}$ with $\gamma_0 =
200$ resulting from a balance
between second-order energy gains and synchrotron losses
\citep{sch85}.  The stochastic acceleration power of radio electrons
is therefore about equal to the observed radio luminosity
$L_{rad}$.

The SSC emission component of these electrons, shown in Fig.\ 1 by the
dashed curve, falls in the sub-keV region below the level of the
quiescent X-ray flux.  However, if magnetic fields $\lesssim 5$ G are
assumed, this SSC contribution would rise and also move to 1 keV, and
could contribute to radiation at that energies. It cannot, however,
explain the entire quiescent flux extending to 10 keV.

The powerful X-ray and NIR flares observed from \sgr are explained in
our model by the onset of instabilities of the accretion flow at
distances of only few $R_{S}$. They result in strong shocks in the
accretion flow and effective acceleration of particles advected with
the flow.  The accelerated particles are then easily taken out of this
region in the wind/jet outflow. Synchrotron emission of electrons
accelerated to $\gamma > 10^{6}$ explains the X-ray flares with very
short variability scales. Furthermore, while propagating
through the first few tens of $R_{S}$, these electrons have sufficient  
time,  $\sim 10^3 \,\rm s$, to cool in the high
$B$ fields there down to $\gamma \lesssim 3\times 10^3$ 
to produce flares in the NIR domain. Self-absorbed flares 
at $\lesssim 100\,\rm
GHz$ detected on $\lesssim 1\,$day timescales after the X-ray flares
\citep{zha04} could be explained by the radiation from these same 
electrons at later stages/larger distances of the outburst in the
`expanding source' scenario.

In Fig.\ 2 we show the flare fluxes expected in this model.
After shock-acceleration close to the BH and injection into
collimated wind outflow with speed $c/2$, electrons propagate through
the inner $r\simeq$10\,-\,30 region on timescale 
$t_{esc}=1200\,\rm s$, after which the radiation
losses drop because of wind expansion and decline of $B$.  The
solid, dashed, and dot-dashed curves show synchrotron and Compton
fluxes produced at times 200\,s, 1200\,s and 1\,hr after the onset of
the flare.  The electron injection time profile is $L_{e.flare}(t)=
L_{0}/ (1+t/t_0)^{2}$, with $t_0 = 720\,\rm s$.  The mean magnetic
field in ADAF outflow could be 
enhanced at the flaring state, so we take $B=25\,\rm G$.  The Compton
fluxes shown (thin curves) clearly demonstrate that no detectable TeV
flares should be expected during powerful X-ray flares. This is in
agreement with the non-detection of TeV flux variations during many
days of observation with the HESS telescopes, whereas the X-ray flares
occur with frequency of $\sim 1$ per day \citep{Chandra}.

\section{Summary and Conclusions}

A model consisting of magnetized coronal ADAF (synchrotron) radio emission
within $\approx 20~R_{\rm S}$ of the GCBH, and of a black-hole plerion
powered by a wind from the ADAF resolves many puzzling observations of
Sgr A$^\ast$.  X-ray flares are synchrotron radiation of electrons
accelerated through first-order Fermi process by shocks within a few
\rs of the GCBH.  Electrons accelerated to $\gamma\gtrsim 10^8$ at the
wind termination shock at $\gtrsim 3\times 10^{16}\,$cm creates the
GCBH plerion.  Its synchrotron X-ray emission has been resolved as an
$\simeq 1.4^{\prime\prime}$ source with Chandra \citep{Chandra}.  The
multi-TeV electrons Compton scatter the radio photons from \sgr and
FIR photons from the dust ring of Sgr A West to produce the TeV
emission.  TeV emission from the GCBH plerion is nearly stationary
because the cooling time of TeV electrons is $\sim 100$ yrs.

A ``jet-ADAF" model for \sgr has been proposed by \citet{fm00} and
\citet{ymf02}. Our model, though also based on energy outflow from the
ADAF, differs greatly.  In particular, the quiescent and flaring X-ray
components are produced at different sites.  The origin of the X-ray
and NIR flares are explained as synchrotron emission of electrons
accelerated during episodes of instabilities very close to the BH, not as
Compton radiation in the SSC scenario of the ``jet-ADAF" model.
Importantly, the TeV flux cannot be easily explained if the observed X-rays 
were due to the SSC mechanism in the vicinity of the GCBH, as
suggested by \citet{fm00}.

Propagation of GeV electrons from the plerion on timescales of
$\gtrsim 10^4 \,\rm yr$ with speeds $\sim 100 \,\rm
km\, s^{-1}$could significantly contribute to the radio synchrotron
flux of Sgr A West. In particular, it could explain the suggested
nonthermal origin of radiation in the bar of Sgr A West \citep{bar}, 
 formed in the outflow direction. 

We also predict that quasi-stationary Compton and bremsstrahlung
fluxes from the pc-scale plerion, coincident with the central parts of
Sgr A West, will be significantly detected and possibly resolved with
GLAST at GeV energies. But the expected highly variable Compton
counterpart of the synchrotron X-ray flares from the GCBH vicinity is
too weak to be detectable with GLAST or HESS.  The synchrotron
extension of the flare emission in the hard X-ray/soft $\gamma$-ray
domain explains the $\simeq 40\,$min episode of a profound increase of
flux detected by INTEGRAL from the direction of
\sgr \citep{bel04}.

The absence of apparent TeV flux variations may suggest a
proton origin for the TeV radiation \citep{an04}. Indeed, proton
and ion acceleration through first- and second-processes in the 
ADAF and through first-order acceleration at the wind termination
shock could make cosmic rays to produce TeV emission through
nuclear $pp$-interactions with $n\sim 10^{3}\,\rm cm^{-3}$ dense 
gas on $pc$ scales, and could form extended TeV 
emission possibly already detected with HESS  
\citep{HESS}.  The hadronic origin of the TeV radiation 
in the ADAF itself in the BH vicinity requires, however, an extremely dense
gas target or extremely large proton powers 
$\gtrsim 10^{39}\,\rm erg\, s^{-1}$.

Finally, we note that the unidentified EGRET source 3EG J1746-2851
towards the GC \citep{EGRET1} is significantly displaced \citep{dh02}
from the direction of the GCBH, and is unlikely to be related to Sgr
A$^\ast$. It is probably emission from a young pulsar, though not with
the ``mouse" PSR J1747-2958 \citep{mc03}. A young pulsar with
$\gamma$-ray properties like Vela but with apparent $\gamma$-ray
power $\approx 10 \times$ larger could have been missed in pulsar
surveys due to the large dispersion measure towards the GC. We suggest
a deeper, higher radio-frequency and X-ray search at the refined 
location of 3EG J1746-2851.

\acknowledgments{We thank I.\ Grenier, S.\ Markoff, R.\ Narayan, P.\ 
Ray, and R.\ Romani for discussions.  
Research of CD and visits of AA to the NRL High
Energy Space Environment Branch are supported by {\it GLAST} Science
Investigation No.\ DPR-S-1563-Y. The work of CD is supported by the
Office of Naval Research.  }

\clearpage

\begin{figure}
\epsscale{1.}
\plotone{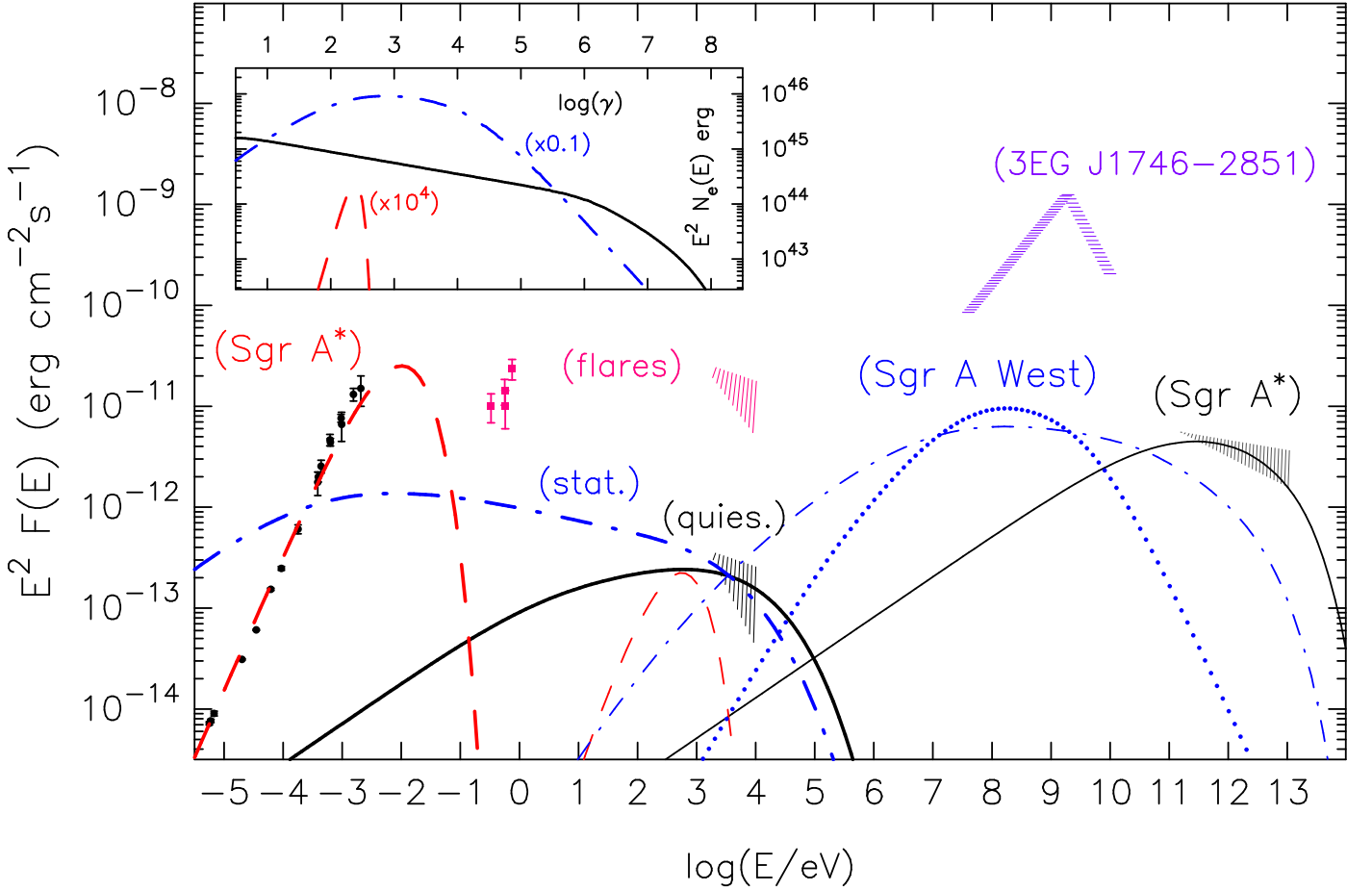}
\caption{Quasi-stationary/quiescent synchrotron ({\it heavy curves})
and Compton ({\it thin curves}) fluxes expected from Sgr
A$^\ast$. Dashed curves show the fluxes from the magnetized ADAF
corona with $B= 10\,\rm G$ and $R\simeq 20\,\rm R_{S}$ due to a
relativistic Maxwellian distribution of electrons with mean
Lorentz-factor $\gamma_{0} = 200$. The solid curves show the fluxes
from the BH plerion with $B_1=90\,\rm \mu G$ at a
distance $R_{pler} \sim 5\times 10^{16}\,\rm cm$; the calculations
assume escape of the flow from the TeV plerion to $R\geq 10^{17}
\,\rm cm $ on timescale $50\,\rm yr$. 
The {\it dot-dashed curves} show the fluxes formed during assumed
$t_{esc} =2\times 10^{4}\,$yr propagation time 
in the plerion at scales $R\sim 0.03$ --
$3\,\rm pc$. The {\it dotted curve} shows the bremsstrahlung from the
central few pcs of Sgr A West with gas density $n_{gas}=10^3\, \rm
cm^{-3}$. The inset shows the electron energy distributions 
formed in those 3 spatial scales. The hatched regions in the X-ray domain
show the fluxes detected by Chandra ($\alpha_X \sim 2.2 - 3$ are
plotted) and XMM ($\alpha_X = 2.5 \pm 0.3$), and the TeV fluxes
detected by HESS (see text for references). The radio data are from
\citet{Z95} and \citet{fal98}. }
\label{f1}
\end{figure}

\begin{figure}
\epsscale{0.95}
\plotone{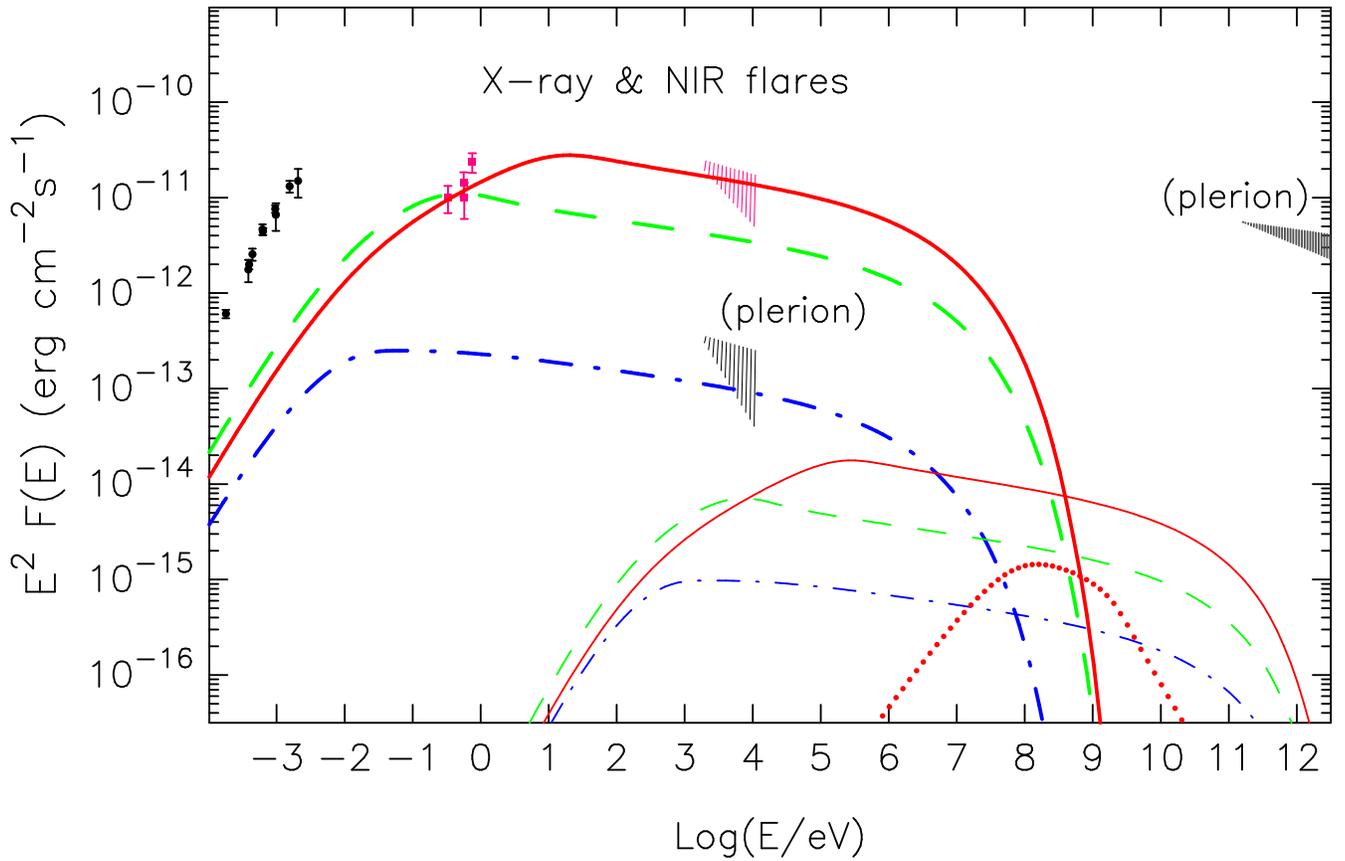}
\caption{ Synchrotron and Compton fluxes 
({\it thick} and {\it thin} curves, respectively) expected during a
flare at 200\,s ({\it solid}), 1200\,s ({\it dashed}), and 3600\,s
({\it dot-dashed} curves) after the onset of the injection of
relativistic electrons with $\alpha_{pler}= 2.2$ and $\gamma_{max}=10^7$,
initial injection power $L_e(0)= 6.3\times 10^{36} \,\rm erg\, s^{-1}$,
and characteristic decline time $t_0 = 720\,\rm s$ (see text).  Escape
of electrons with the conical wind during $t_{esc} = 1200\,\rm s$ from the
central $R\leq 20 \,\rm R_{S}=1.8\times 10^{13}\,\rm cm$ region of the
ADAF is supposed. The dotted curve shows the expected 
 bremsstrahlung $\gamma$ rays at $t=200\,\rm s$, calculated for 
plasma density of $10^8\,\rm cm^{-3}$ in the ADAF.  }
\label{f2}
\end{figure}

\end{document}